\documentclass[twocolumn,amsmath,amssymb,nofootinbib,superscriptaddress,longbibliography,floatfix]{revtex4-1}

\usepackage{amsmath,amsfonts}
\usepackage{algorithmic}
\usepackage{algorithm}
\usepackage{array}
\usepackage[caption=false,font=normalsize,labelfont=sf,textfont=sf]{subfig}
\usepackage{textcomp}
\usepackage{url}
\usepackage{verbatim}
\usepackage{graphicx}
\usepackage{textcomp}
\usepackage[english]{babel}

\newcommand{\oncircuit}[1]{(\texttt{#1}\kern-0.65pt)}

\usepackage{hyperref}
\usepackage{etoolbox}

\makeatletter
\patchcmd{\@outputpage@head}{\@ifx{\LS@rot\@undefined}{}{\LS@rot}}{}{}{}
\makeatother

\usepackage{pdfpages}

\begin{document}

\preprint{APS/123-QED}

\title{Swarmodroid \& AMPy: Reconfigurable Bristle-Bots and Software Package for Robotic Active Matter Studies}

\author{Alexey~A.~Dmitriev}
\thanks{Alexey A. Dmitriev and Vadim A. Porvatov contributed equally to this work}
\affiliation{School of Physics and Engineering, ITMO University, Saint Petersburg 197101, Russia}

\author{Vadim~A.~Porvatov}
\thanks{Alexey A. Dmitriev and Vadim A. Porvatov contributed equally to this work}
\affiliation{School of Physics and Engineering, ITMO University, Saint Petersburg 197101, Russia}

\author{Alina~D.~Rozenblit}
\affiliation{School of Physics and Engineering, ITMO University, Saint Petersburg 197101, Russia}

\author{Mikhail~K.~Buzakov}
\affiliation{School of Physics and Engineering, ITMO University, Saint Petersburg 197101, Russia}

\author{Anastasia~A.~Molodtsova}
\affiliation{School of Physics and Engineering, ITMO University, Saint Petersburg 197101, Russia}

\author{Daria~V.~Sennikova}
\affiliation{School of Physics and Engineering, ITMO University, Saint Petersburg 197101, Russia}

\author{Vyacheslav~A.~Smirnov}
\affiliation{School of Physics and Engineering, ITMO University, Saint Petersburg 197101, Russia}

\author{Oleg~I.~Burmistrov}
\affiliation{School of Physics and Engineering, ITMO University, Saint Petersburg 197101, Russia}

\author{Timur~I.~Karimov}
\affiliation{Department of Computer-Aided Design, Saint Petersburg Electrotechnical University "LETI", Saint Petersburg 197376, Russia}

\author{Ekaterina~M.~Puhtina}
\affiliation{School of Physics and Engineering, ITMO University, Saint Petersburg 197101, Russia}

\author{Nikita~A.~Olekhno}
 \email{nikita.olekhno@metalab.ifmo.ru}
\affiliation{School of Physics and Engineering, ITMO University, Saint Petersburg 197101, Russia}

\date{\today}

\begin{abstract}
Large assemblies of extremely simple robots capable only of basic motion activities (like propelling forward or self-rotating) are often applied to study swarming behavior or implement various phenomena characteristic of active matter composed of non-equilibrium particles that convert their energy to a directed motion. As a result, a great abundance of compact swarm robots have been developed. The simplest are bristle-bots that self-propel via converting their vibration with the help of elastic bristles. However, many platforms are optimized for a certain class of studies, are not always made open-source, or have limited customization potential. To address these issues, we develop the open-source Swarmodroid~1.0 platform based on bristle-bots with reconfigurable 3D printed bodies and simple electronics that possess external control of motion velocity and demonstrate basic capabilities of trajectory adjustment. Then, we perform a detailed analysis of individual Swarmodroids' motion characteristics and their kinematics. In addition, we introduce the AMPy software package in Python that features OpenCV-based extraction of robotic swarm kinematics accompanied by the evaluation of key physical quantities describing the collective dynamics. Finally, we discuss potential applications as well as further directions for fundamental studies and Swarmodroid~1.0 platform development.
\end{abstract}

\maketitle

\section{Introduction}
\label{sec:Introduction}

Emergent phenomena in large assemblies of moving agents that are guided statistically rather than controlled directly are considered as one of the key topics at the intersection of physics and robotics. From the physics perspective, large swarms of moving particles form a class of non-equilibrium soft matter systems known as active matter~\cite{2013_Marchetti, 2012_Vicsek}, often implemented with macroscopic artificial agents such as robots~\cite{2024_Ning}.

{In swarm robotics, many modern approaches towards the control of a swarm rely on various self-organization and biomimetic effects instead of centralized control. Recent examples addressing programmable robots include an emulation of tissue morphogenesis in swarms of robots that implement local algorithms~\cite{2018_Slavkov}, machine learning-based control of swarms of simple robots~\cite{2022_Shen, 2023_Zion}, or even the analog of a self-organized nervous system for swarm coordination~\cite{2024_Zhu}, to name a few. Moreover, various control approaches that utilize analogies between robots and particles whose behavior is governed by specific physics are considered, including statistical-based control~\cite{2019_Li}, the implementation of hydrodynamic equations~\cite{2023_Shibahara}, an engineered rattling~\cite{2019_Savoie, 2021_Chvykov}, and cohesive interactions between robots~\cite{2021_Li}.}

{To better contextualize our work, we now focus on recent examples when various biomimetic or self-organization control strategies emerge in swarms of simple, non-programmable bristle-bots converting vibration of a motor into a directed motion with the help of elastic bristles~\cite{2013_Giomi} and interacting with each other mostly through collisions. First, such bristle-bots were placed inside a flexible boundary to form motile superstructures~\cite{2018_Deblais, 2021_Boudet} capable of adaptive transport through narrow regions and performing simple actions, such as removal of debris from a test area. An increased transport adaptability was recently demonstrated in~\cite{2024_Xi}, with bristle-bots linked by a flexible beam that successfully travel through a maze. Moreover, biomimetic formation of large structures by bristle-bots with specifically engineered body shapes has been considered~\cite{2016_Andreen, 2024_Pan}. Finally, bristle-bots have been deployed in heterogeneous robotic collectives for structural health monitoring~\cite{2025_Fath}.} Further miniaturization of non-programmable robots looks promising for the realization of experiment automation at millimeter scales~\cite{2020_Yu}, removal of microplastics~\cite{2022_Fu, 2022_Li}, and even medical applications of particle swarms~\cite{2022_Fu} at the microscale.

The first compact swarm robotic platforms employed two-wheeled robots, usually incorporating infrared sensors and emitters that facilitate the implementation of interactions between robots and obstacle avoidance. Moreover, such platforms often support the installation of additional devices, including a video camera (e-puck2, Alice)~\cite{gtronic_e_Puck, 2000_Caprari_Alice}, accelerometers (Elisa-3, e-puck2)~\cite{elisa3, gtronic_e_Puck}, microphones (AMiR, e-puck2)~\cite{2009_Arvin_AMIR, gtronic_e_Puck}, ultrasound sensors (Colias)~\cite{2014_Arvin_Colias}, and RF modules (Elisa-3, Alice)~\cite{elisa3, 2000_Caprari_Alice}. Specifically, Elisa-3 is an open-source platform~\cite{elisa3} compatible with Arduino\textregistered\ that incorporates a wide range of sensors, employing robots that can be piloted to the charger station and automatically self-charge~\cite{elisa3}. Swarms of four to $25$ Elisa-3 robots have been used for the development of distributed algorithms for dynamic task execution based on RF communication between robots~\cite{2016_deMendonca}. Colias~\cite{2014_Arvin_Colias}, AMiR~\cite{2009_Arvin_AMIR} and Jasmine~\cite{jasmine} are open-hardware robots that are frequently used in the experimental realization of the BEECLUST algorithm (inspired by the collective behavior of honeybees)~\cite{2009_Schmickl}. The lightweight Alice robotic platform~\cite{2000_Caprari_Alice} stands out because it {features the most compact robots among all previously mentioned and has} a battery capacity that allows them to work for $10$~hours until recharge. As a particular example, a swarm of $20$ Alice robots emulating cockroach aggregation was considered~\cite{2008_Garnier}. Finally, e-puck2~\cite{gtronic_e_Puck} represents the most sophisticated (and most expensive) platform among the considered ones that has an extensive set of sensors. Although studies involving large swarms of e-puck2 robots are unlikely to appear due to the high price of a single robot, even a small number of e-puck2 robots successfully demonstrate occlusion-based cargo transport that requires visual recognition of an object and a target without any communication between the robots~\cite{2015_Chen}.

However, the relatively high prices of the considered wheeled robots limit their affordable number in the swarm, and the experiments in most of the mentioned papers are carried out with systems containing only three to $20$ robots. As a result, many experimental studies of robotic swarm physics employ cheaper bristle-bots that propel by converting the oscillations of a vibration motor to a directed motion with the help of flexible bristles. The simplest commercial bristle-bots are HEXBUGs\textregistered~\cite{Hexbug} that are distributed as toys. Examples of their application include a study of boundary-controlled swarm dynamics~\cite{2018_Deblais, 2021_Boudet}, the emulation of traffic jams~\cite{2019_Barois} and financial price dynamics~\cite{2020_Patterson}, experimental studies of polarized wall currents of self-propelled particles~\cite{2020_Barois}, the analysis of a single robot in a parabolic potential~\cite{2019_Dauchot}, statistical physics of such non-equilibrium swarms~\cite{2022_Boudet, 2023_Chen}, and educational analogies~\cite{2022_DiBari}. However, HEXBUGs\textregistered\ cannot be turned on and off simultaneously using a remote control and need to be placed in the system one by one after being manually turned on, which may affect the physics of the swarm. Moreover, their motion characteristics (governed by the shape of bristles) and motor vibration patterns cannot be controlled as well. Thus, the only degree of freedom left to address new physics is to change the shape of HEXBUGs\textregistered\ bodies or append them with additional elements. For example, HEXBUGs\textregistered\ have been supplied with magnets to demonstrate a magnetotaxis inspired by biological systems~\cite{2021_Sepulveda} and have been turned into chiral self-rotating matter by merging two HEXBUG\textregistered\ bristle-bots together to study the emergence of robust edge currents~\cite{2020_Yang}.

When more specific functionalities are required, custom hardware platforms are developed. Such bristle-bots range from rather simple BBots applied to study the swirling and swarming behavior in Ref.~\cite{2013_Giomi} to more advanced BOBbots {that incorporate magnets} to introduce the attraction between robots {along with} wireless chargers to simplify their maintenance~\cite{2021_Li}. Cooperative transport has been demonstrated in custom-built bristle bots, which allow transporting a load that is too heavy for a single robot to move~\cite{2025_Arbel}. Miniature rotating Magbots with a diameter of $2$~cm, equipped with up to six magnets {and having their} vibration intensity controlled by photoresistive light sensors demonstrate a tunable transition between robot-like movement and matter-like properties~\cite{2024_Wang}. There are also several unusual designs ranging from an extra-small $5$~milligram bristle-bot~\cite{2019_Kim} and magnetic-field driven bristle-bots~\cite{2023_Supik} to a bristless SurferBot, a vibrobot capable of moving on the surface of water~\cite{2022_Rhee_Surferbot}.

Another widely used robots are Kilobots~\cite{2012_Rubenstein_Kilobot} which represent an example of a polished and versatile platform designed to carry out scientific experiments. Being able to communicate through optical channels with the help of light-emitting diodes and supplied with two vibration motors and a programmable microcontroller, Kilobots were applied, for example, to demonstrate self-organization in predefined shapes following local algorithms~\cite{2014_Rubenstein} and emulate tissue morphogenesis~\cite{2018_Slavkov}. However, by their design, Kilobots move slowly and via discrete steps, thus being more suitable for hardware implementations of various cellular automata and other mathematical models rather than to study physics governed by mechanical interactions between robots such as collisions at high speeds and the formation of force chains. To our knowledge, there is a single example of faster-moving Kilobots ($5$~cm/s instead of $0.5$~cm/s) supplied with additional 3D-printed tripods~\cite{2023_Zion}.

\begin{figure*}[tbp]
    \centering
    \includegraphics[width=\textwidth]{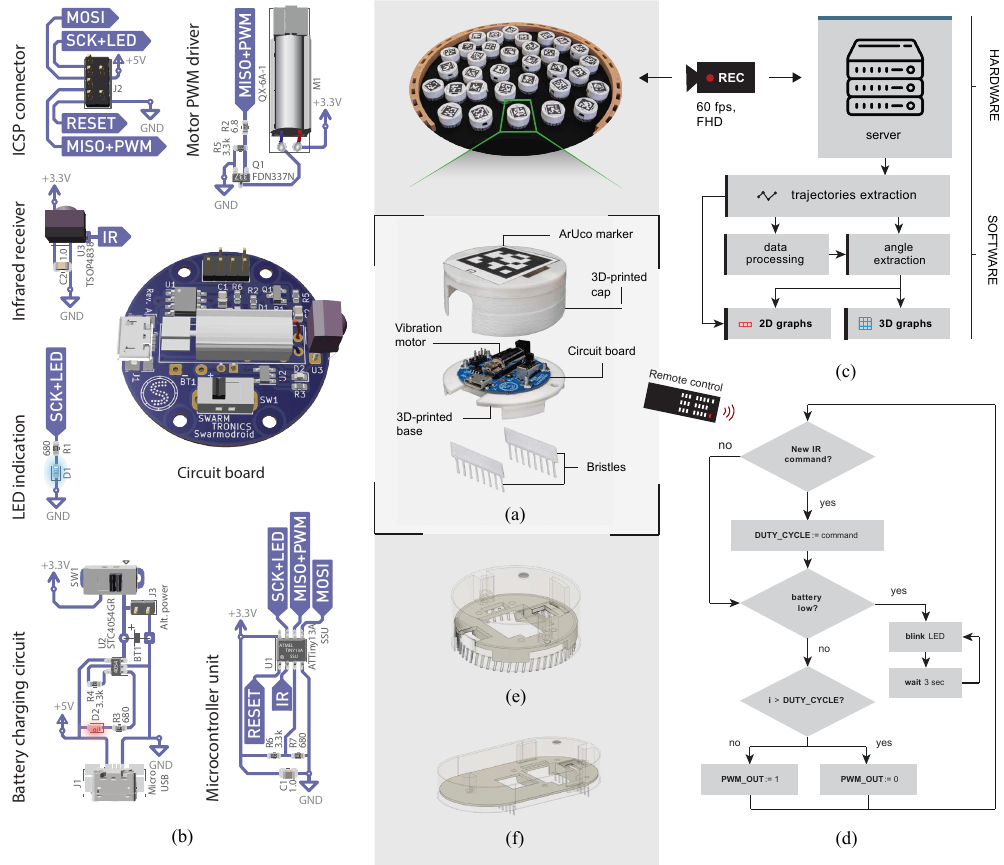}
    \caption{Schematics of the Swarmodroid platform. (a) Robotic swarm confined in a circular-shaped barrier (top) and the burst diagram of a single robot (bottom). The robot consists of a 3D printed cap, base, bristles, and a printed circuit board with a vibration motor. Individual markers (ArUco or AprilTag) are placed on the top surfaces of robots. (b) Diagram of the control circuit showing the labeled key blocks of the circuit along with its render (in the center). (c) Processing software diagram. The motion of the robots is captured with the help of an HD camera, and the locations of the markers are extracted via the OpenCV library. Then, various quantities characterizing single-robot dynamics as well as collective behavior are evaluated. (d) Diagram of the Swarmodroid firmware executed in the ATTiny13 microcontroller on the circuit board. The robot checks the presence of an IR remote controller command and, if present, adjusts its motion velocity. In addition, the battery level is checked and displayed via the LEDs. (e,f) Different designs of 3D printed bodies corresponding (e) to cylindrical self-rotating Type-I Swarmodroids and (f) to elongated self-propelled Type-II Swarmodroids. Both types are assembled with the same circuit board from Panel (b).}
    \label{fig:Platform}
\end{figure*}

In this paper, we introduce Swarmodroid~1.0 bristle-bots that are USB-rechargeable, feature remotely controlled motion velocity, reconfigurable plastic bodies, and variable motion patterns. Our hardware platform is supplied with a software counterpart, the open-source AMPy software library in Python capable of extracting and visualizing swarm kinematics. Both parts of the platform are distributed under GPL~v3 license~\cite{GPLv3}. The production of Swarmodroids requires the use of a single-layer printed circuit board, additive manufacturing, and widely spread, affordable components, thus being accessible both for low-cost tests and for the implementation of swarms with large numbers of robots.

The paper is organized as follows. In Section~\ref{sec:Design}, we describe the structure of the experimental setup as well as the basic building blocks of individual Swarmodroids. Then, in Section~\ref{sec:Circuit}, we focus on the Swarmodroid printed circuit board and its programming. Section~\ref{sec:Software} introduces the processing software within the AMPy package. Section~\ref{sec:Individual} addresses the kinematic characteristics of individual Swarmodroids in different regimes averaged over several robots. Finally, potential applications and directions for future development are summarized in Section~\ref{sec:Outlook}.

\section{Robot design}
\label{sec:Design}

An experimental setup to study collective effects in robotic swarms typically includes a barrier that surrounds some area in which robots move and a set of robots, Figure~\ref{fig:Platform}(a). In our case, the setup is supplied with a Sony ZV-E10 HD camera that captures the motion of robots. Then, the locations of the markers placed in the center of the top surface of each robot are extracted with the help of OpenCV library and processed by the introduced AMPy software package, Figure~\ref{fig:Platform}(c).

The Swarmodroid body consists of several plastic elements, including a cap with an ArUco or an AprilTag marker, a base to which the printed circuit board (PCB) is attached, and bristles that convert the vibration of the robot motor to directed motion\footnote{https://github.com/swarmtronics/swarmodroid.pcb}, see Figure~\ref{fig:Platform}(a). All these parts are produced using fused deposition modeling (FDM) technology using Flying Bear Ghost 5 3D printers with nozzle diameter $0.4$~mm. As the printing material, PLA plastic with a melting point around $200^{\circ}$C was chosen. In the following, we address in detail two particular variations of Swarmodroids.

\paragraph*{Circular-shaped self-rotating robots (Type-I)} For such robots, the base has the form of a circular plate with diameter of $46$~mm and thickness of $1.5$~mm, Figure~\ref{fig:Platform}(e). The base features a section for a rechargeable battery in the center, round holes for the screws fastening the circuit board and the cap, and several rectangular holes matching the elements of PCB such as a USB-port for charging. The top surface of the base is totally flat, while the bottom surface contains protrusions for the attachment of bristles and a battery. The bristles can be attached in two configurations, allowing us to implement the self-propelled and self-rotating types of motion, respectively. In the first case shown in the inset of Figure~\ref{fig:Platform}(a), two lines of bristles inclined at the angle $10^\circ$ (counting from normal to surface) are attached to straight grooves, while in the second case the bristles with the same angle of inclination are attached in a closed line along the edge of the base, as in Figure~\ref{fig:Platform}(e). In such a case, the clockwise or counterclockwise rotation of the robots is defined by the slope direction. For both configurations, a single bristle has dimensions of $7.5 \times 0.8 \times 0.4$~mm. {The selected inclination angle corresponds to the highest angular velocity of robots rotation according to the measurements of Ref.~\cite{2021_Porvatov}.}

The cap has a cylindrical shape with a diameter ${d=48.7}$~mm, a height of $19.2$~mm, and a wall thickness of $0.6$~mm. The top surface of the cap features technological holes providing access to the switcher allowing us to manually turn the robot on and off, and LEDs indicating the state of the robot. In addition, there are two pillars inside the cap containing a section for a nut and a hole for a screw, and a single additional supporting pillar attached to the base via the corresponding notch. The height of the pillars is $10.1$~mm, which corresponds to the height of the top surface of the cap relative to the base. The rest height of the cap's side surface partially covers the bristles in order to increase the stability of the robot. The aperture on the side surface of the cap allows charging robots without disassembling them.

In the bristle-bot design, flexible bristles play a crucial role by converting the vibration of a motor to the motion of the robot. Despite the fact that the bristles are made of rigid PLA plastic (to simplify robot production by using the same material for all parts), they are still flexible enough due to their small thickness of $0.4$~mm. As shown in Ref.~\cite{2021_Porvatov}, such PLA bristles even outperform the bristles made of BFlex resin-like material when implementing self-rotating robots. The inclination angle of the bristles equal to $10^\circ$ is chosen as an optimal value for the PLA bristles according to the reference above, as well as taking into account the results of work~\cite{2013_Giomi}.

\paragraph*{oval-shaped self-propelled robots (Type-II)} The bases of such robots have a racetrack-like geometry with axes $82.6$~mm and $45$~mm, respectively. The location of PCB is shifted from the geometrical center, yet the circuit board is attached in the same manner as for Type-I Swarmodroids. However, the size of the base and the cap, which is larger compared to self-rotating design, allows to implement the assembly of these two parts via neodymium magnets instead of screws. We utilize magnets with a cylindrical shape with a diameter of $3$~mm and a height of $2$~mm for this purpose. The magnets are located at the opposite edges of the robot on its larger axis and are placed in the slots with a matching geometry in the base and cap. The bottom side of the base has three straight grooves inclined at an angle $10^\circ$ measured from the normal to the surface at which the bristles are placed. However, to achieve an efficient self-propelled motion, it is sufficient to use just two sets of bristles placed at the outermost grooves.

In the following, we study the properties of Type-I and Type-II Swarmodroids and perform two sets of experiments with large robotic collectives, one with the self-rotating Type-I robots and the other one with the self-propelling Type-II. However, there are unlimited possibilities to design various robot shapes while using the same circuit board and maintaining compatibility with recognition software.

\section{Robot circuitry}
\label{sec:Circuit}

The circuit diagram is shown in Fig.~\ref{fig:Platform}(b). It can be divided into the following structural blocks: (i) battery and a charging circuit, (ii) LED indication, (iii) motor pulse-width modulation (PWM) driver, (iv) infrared receiver, (v) microcontroller unit (MCU), and (vi) in-circuit serial programming (ICSP) connector, all located at (vii) the printed circuit board (PCB).

\paragraph*{(i) Battery and charging circuit}
All electric components are powered by a Robiton LP601120 100~mAh lithium-ion polymer battery~\oncircuit{BT1}; hereafter, labels in brackets denote the corresponding parts in Fig.~\ref{fig:Platform}(b). The SS12D07 battery disconnect switch \oncircuit{SW1} prevents the circuit from draining the battery while the robot is inactive. Battery charging can be performed from any $5$~V\,DC source that can supply a $300$~mA current, through a Micro USB connector~\oncircuit{J1}, or alternatively through the ICSP connector~\oncircuit{J2}. The actual charging current and voltage delivered to the battery cell are controlled by the charge control circuit STMicroelectronics STC4054GR~\oncircuit{U2}, which limits the charging current to $300$~mA during the constant-current charging phase and the charging voltage to $4.2$~V during the constant-voltage charging phase. Additionally, a footprint for a PLS2-2 pin header~\oncircuit{J3} is provided to connect alternative power sources (such as laboratory power sources, wireless charging coils, etc.), but the pin header itself is not installed if a battery is used.

\paragraph*{(ii) LED indication}
The LED indication consists of two 0603 surface-mount LEDs~\oncircuit{D1} and~\oncircuit{D2} with current-limiting resistors~\oncircuit{R1} and~\oncircuit{R2}. The LED~\oncircuit{D1} serves as a general-purpose indication and is switched by the MCU. It displays the following signals: (a) shining constantly -- robot is running; (b) turned off -- robot switched off; (c) briefly turned off -- receiving a command from IR remote control; (d) blinking one, two or three times -- battery level 30\%, 60\% or 100\% respectively; (d) blinking briefly once every three seconds -- battery level below critical, need to charge the robot immediately to avoid battery overdischarge. The LED~\oncircuit{D2} is driven by the charge controller~\oncircuit{U2} and only has two states: (a) shining -- the battery is now charged, (b) turned off -- charge finished (charging current fell below 30~mA) or charger not connected.

\paragraph*{(iii) Motor PWM driver}
The robot is actuated by the vibration motor QX~Motor QX-6A-1~\oncircuit{M1}. Different motion velocities are implemented by limiting the average motor power to the selected percentage using an Onsemi FDN337N n-MOSFET switch~\oncircuit{Q1} that is driven with a pulse-width-modulated signal. The gate of transistor~\oncircuit{Q1} is pulled down to the source by the resistor~\oncircuit{R5} to prevent spontaneous opening. The resistor~\oncircuit{R2} acts as a gate current limiter. The PWM signal has a frequency of approximately $70$~Hz and $3.3$~V CMOS logic level. The PWM duty cycle allows $256$ steps, from $0\%$ (completely off) to $100\%$ (completely on). In the limiting cases of the duty cycle equal to $0\%$ and $100\%$, the PWM is turned off and a constant gate voltage is provided instead.

\paragraph*{(iv) Infrared receiver}
To capture commands sent by an infrared remote control device, Vishay TSOP4838 infrared receiver~\oncircuit{U3} is used. The circuit is designed to accept commands transferred by the NEC infrared protocol~\cite{2022_Gorodechny}. The receiver~\oncircuit{U3} accepts a sequence of $38$~kHz pulse bursts of the infrared signal and sends a $3.3$~V logic level signal to the MCU according to the following rule: logical low if a pulse burst is being received, logical high otherwise. The resulting logical pulse sequence is a portion of pulse-period modulated data, which is software-decoded by the MCU. A $1\,\text{\textmu F}$ filtering capacitor~\oncircuit{C2} is installed near the receiver~\oncircuit{U3} to isolate it from switching noise. {According to the NEC infrared protocol specifications, the delay between signal receiving and the start of robots motion is $67.5$~ms~\cite{2022_Gorodechny}.}

\paragraph*{(v) Microcontroller unit}
The entire Swarmodroid circuit, excluding the charging subsystem, is controlled by the Microchip ATTiny13A-SSU AVR microcontroller~\oncircuit{U1}. It performs the following functions: generation of the PWM signal that drives the \oncircuit{Q1} gate, decoding the pulse sequences sent by the IR receiver~\oncircuit{U3}, general-purpose indication via the LED~\oncircuit{D1}, as well as battery voltage supervision by measuring the voltage on the resistor divider~\oncircuit{R6}-\oncircuit{R7} to prevent overdischarge. A $1\,\text{\textmu F}$ filtering capacitor~\oncircuit{C1} is installed near the MCU to reduce its sensitivity to switching noise.

The MCU firmware~\footnote{https://github.com/swarmtronics/swarmodroid.firmware} performs the following actions. First, as soon as the robot is turned on, a self-test is performed to ensure that the battery voltage is above the critical level (approximately $3.3$~V). If it is below this threshold, the robot enters the power-saving mode. Otherwise, the measured battery idle voltage is indicated by blinking the LED~\oncircuit{D1} one time for a low charge level, two times for a medium level, and three times for a full charge, respectively. After that, the motor is tested by turning it on for a time period of $50$~ms, and the LED is lit continuously to indicate that the robot is ready. At this stage, the main code enters an infinite waiting loop (the main loop), which is terminated when the battery voltage drops below the critical level. Upon such an event, the robot enters the power saving mode.

\begin{figure*}
    \centering
    \includegraphics[width=\textwidth]{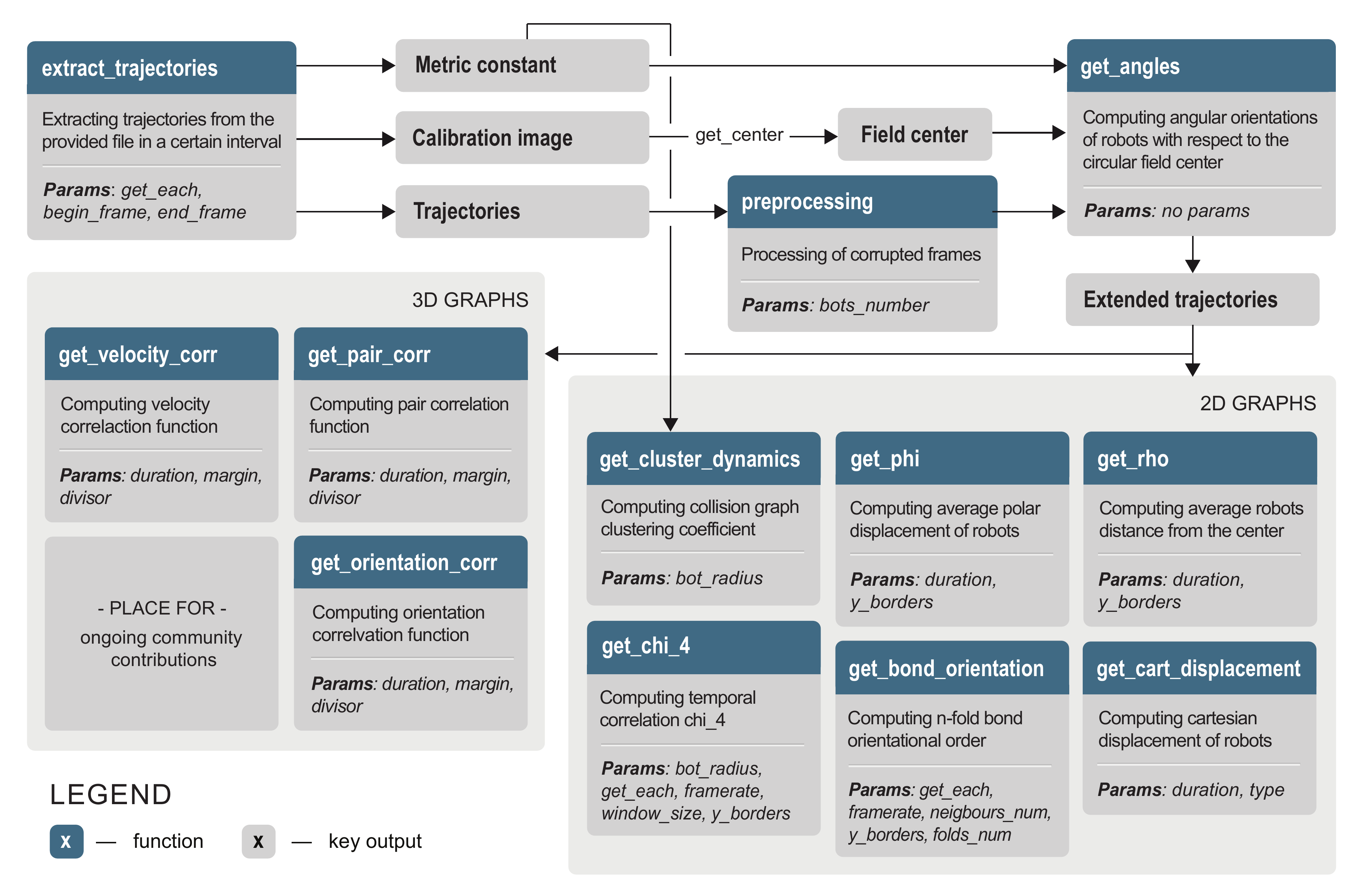}
    \label{fig:Software}
    \caption{Software diagram illustrating the processing pipeline of experimental data. Each code block includes the name of the corresponding function, a brief description of its content, and a list of input parameters. }
\end{figure*}

While the main loop is running, the $8$-bit timer/counter of the MCU (clocked at 18.75~kHz) is used simultaneously to generate the PWM signal, measure the pulse widths to demodulate the signal from the IR receiver, and to trigger periodic battery voltage checks. For a timing diagram, see Supplementary Materials~\cite{Supplementary}. In the main loop itself, no other action is performed except for waiting for an interrupt event, which is caused by an incoming command accepted by the IR receiver~\oncircuit{U3}. The commands are received as pulse sequences consisting of 32 bits encoded with the pulse-period modulation as defined by the NEC protocol (we are using its variant with a $16$-bit address). Each logical level change on the~\oncircuit{U3} output generates an interrupt event~\cite{2010_Attiny13}. The demodulation is performed in the corresponding interrupt service routine of the MCU by measuring the time intervals between the falling edges of the pulses, using the $8$-bit timer/counter. If the command received is valid according to the NEC protocol and its address part is equal to the hard-coded address constant of the robot, the corresponding action is taken. The basic use case is pairing the robot with a TV remote control and using the digit keys to control the robot -- in this case, the actions are to set the PWM duty cycle to $\text{Duty~Cycle} = \text{Digit} \cdot 10\%$. The power button is used to set the duty cycle to zero, and the $0$ button -- to $100\%$. During the incoming IR pulse sequence, the LED~\oncircuit{D1} is turned off to indicate that a command is currently received.

Finally, as soon as the battery voltage falls below the critical level, the robot is forced into power saving mode. In this mode, all interrupts are disabled, i.e., the robot is rendered unresponsive to any commands; the motor is turned off by sending a constant logical low to the gate~\oncircuit{Q1}. In addition, the LED~\oncircuit{D1} is turned off and briefly blinked every three seconds to indicate that the robot needs to be charged.

The flow chart as well as the complete description of the robot firmware is provided in the Supplementary Materials~\cite{Supplementary}.

\paragraph*{(vi) ICSP Connector}
The MCU can be reprogrammed by the serial peripheral interface (SPI) through an ICSP connector~\oncircuit{J2} (the power switch \oncircuit{SW1} must be in the closed position during programming). Note that two of the~\oncircuit{U1} pins share multiple functions: pin~$6$ drives the gate~\oncircuit{Q1} and doubles as the SPI MISO pin, while pin~$7$ is used for the indication by the LED~\oncircuit{D1} and doubles as the SPI SCK pin. This approach allows to easily verify that a robot is actually being programmed: in a valid programming procedure, the motor vibrates, and the LED rapidly blinks.

\paragraph*{(vii) Printed circuit board}
All electric components are mounted on a two-sided printed circuit board (PCB) in the shape of a disk with a diameter $35$~mm, made on a $0.8$~mm thick FR4 dielectric substrate with $18~\text{\textmu m}$ copper layers. The PCB is mounted bottom side to base top using four DIN-7985/ISO-7045 M2$\times$6 screws and DIN-439/ISO-4035 M2 nuts. For these screws, four mounting holes of $2.2$~mm diameter are provided in the PCB. In turn, all electronic components aside from the battery are mounted on the top side of the PCB. The only component on the bottom side is the battery, which is fitted into a specially designed notch in the base and connects to the PCB using wires.

\section{AMPy experiment processing software}
\label{sec:Software}

\begin{figure*}
    \centering
    \includegraphics[width=\textwidth]{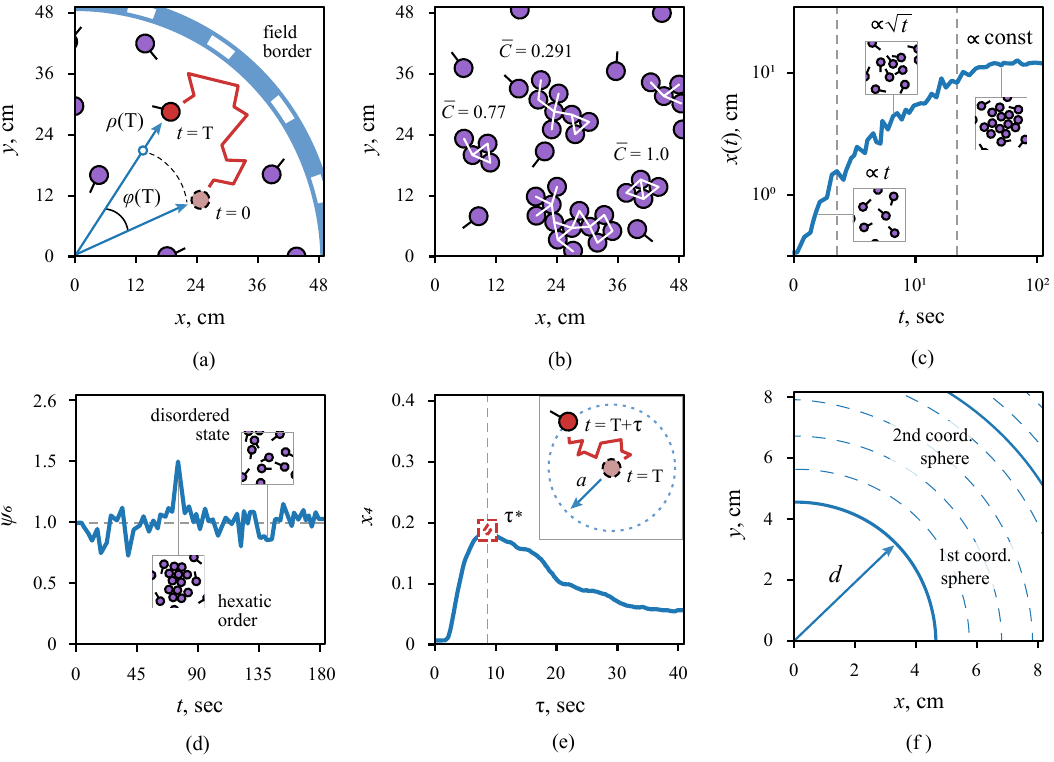}
    \label{fig:char_demonstration}
    \caption{Illustration of different quantities extracted by the software. (a) Several robots (purple circles) placed in a circular barrier. The trajectory of a selected robot between the timestamps $t=0$ and $t=T$ is shown with a solid red line, the line segments near the circles denote the velocity direction for each robot, $\rho(t)$ and $\varphi(t)$ are the radius and polar angle of the robot in polar coordinates centered at the center of the barrier, respectively. (b) Several clusters of touching robots. Force chains are shown with solid white lines. The values of the average clustering coefficient $\overline{C}$ shown near the corresponding clusters are evaluated with Eq.~\eqref{eq:Clustering}. (c) Root mean square displacement $x(t)$ Eq.~\eqref{eq:RMS} schematically demonstrating the transition between ballistic (a linear region $x \propto t$), diffusive (a square root dependency $x \propto t^{1/2}$) and jamming (a saturated region $x \propto \text{const}$) behaviors. The insets show robotic swarm patterns with various densities characteristic of the corresponding motion types. (d) Sixfold index $\psi_{6}$ Eq.~\eqref{eq:Sixfold} demonstrating the transition between a disordered phase (low values) and a hexatic order (peak). The insets demonstrate the characteristic geometries of the system with and without hexatic order. (e) Spatio-temporal correlation parameter $\tau^{*}$ Eq.~\eqref{eq:tau} for a robot with a characteristic localization time close to $10$~s. The inset demonstrates the robot's trajectory between timestamps $t=T$ and $t=T+\tau$. (f) Sketch of the two-dimensional pair correlation function Eq.~\eqref{eq:Correlation} for Type-I Swarmodroids showing characteristic circles at the distances of the robot diameter $d=46$~mm (the first coordination sphere) and two robot diameters (the second coordination sphere) along with the intermediate circles corresponding to other characteristic configurations of robots.  }
\end{figure*}

In addition to the hardware part of the platform, we introduce the AMPy package for video data processing~\footnote{https://github.com/swarmtronics/ampy}. The code is written in Python and evaluates various physical quantities, providing insight into the collective properties of the swarm, Figure~\ref{fig:Software}.

First, the package allows one to extract the coordinates and orientations of individual robots by recognizing the ArUco/AprilTag markers placed on top of each robot, Figure~\ref{fig:Platform}(b). As the evaluation of statistical characteristics relies on the coordinates of the center of area filled with robots, we implemented a special widget allowing to obtain such a point by detecting four auxiliary markers placed at the barrier. In order to eliminate any video distortions that reduce the visibility of the markers, we linearly interpolate missing points during the preprocessing stage. After such an extension of initial trajectories, we determine the robots' orientations with the help of known positions of the markers, Figure~\ref{fig:char_demonstration}(a). The final part of the pipeline allows one to extract different types of motion characteristics of robotic swarms, including time dependencies of robot displacements, correlation functions, and collision graphs, as described further in the text.

\subsection{Collision graph statistics}

\textit{Average clustering coefficient}~\cite{2020_Bindgen}
is the parameter quantifying the density of the force chains induced by robots collision:
\begin{equation}
   \overline{C}=\frac{1}{N} \sum_{i=1}^{N} C_{i},
   \label{eq:Clustering}
\end{equation}
where $C_i$ is the local clustering coefficient of i'th node evaluated as follows:
\begin{equation}
   C_{i}=\frac{1}{k_{i}\left(k_{i}-1\right)} \sum_{j,k} A_{ij}A_{jk}A_{ki},
\end{equation}
where $k_{i}=\sum_{j} A_{ij}$, and $\hat{A}$ denotes the adjacency matrix of the collision graph. As seen in Figure~\ref{fig:char_demonstration}(b), higher values of $\overline{C}$ correspond to a greater number of contacts between touching robots within a cluster, i.e., to more rigid and densely packed clusters.

\subsection{Displacement-based statistics}

\textit{Average displacement}~\cite{2020_Breoni} allows characterizing the motion type of robots, as demonstrated in Figure~\ref{fig:char_demonstration}(c), and in Cartesian coordinates reads 
\begin{equation}
   x(t)=\frac{1}{N} \sum_{i=1}^{N} \sqrt{\left(x_{0}^{(i)}-x_{1}^{(i)}(t)\right)^{2}+\left(y_{0}^{(i)}-y_{1}^{(i)}(t)\right)^{2}},
   \label{eq:RMS}
\end{equation}
where $(x_{0}^{(i)}, y_{0}^{(i)})$ is the initial position of the robot corresponding to the $i$'th trajectory and $(x_{1}^{(i)}(t), y_{1}^{(i)}(t))$ is the position of the same robot at the moment $t$. For sparse systems characterized by ballistic motion, the robots move freely between rare collisions, and $x(t)$ features a linear time dependence, see the first region in Figure~\ref{fig:char_demonstration}(c). At higher densities, the robots collide frequently and change their direction of motion, which results in diffusive dynamics characteristic of liquids; see the intermediate region in Figure~\ref{fig:char_demonstration}(c). In this case, $x(t)$ demonstrates a square root dependence on $t$. Finally, at very high densities, the robots form a rigid cluster and slightly fluctuate near their typical locations, with $x(t) \approx \text{const}$, as shown in the rightmost region of Figure~\ref{fig:char_demonstration}(c).

For polar coordinates, we introduce the $\rho$ parameter describing the average distance of robots from their initial positions with respect to the area center:
\begin{equation} 
   \rho(t)=\frac{1}{N}\sum_{i=1}^{N} \left(\rho_{0}^{(i)}-\rho_{1}^{(i)}(t)\right),
   \label{eq:rho}
\end{equation}
where $\rho_{0}^{(i)}$ is the distance between the center of an area and the given robot at the moment $t=0$ while $\rho_{1}^{(i)}(t)$ is the distance at the moment $t$. The parameter $\phi$ captures the dynamics of polar angle displacement:
\begin{equation}
   \varphi(t)=\frac{1}{N}\sum_{i=1}^{N} \left(\phi_{0}^{(i)}-\phi_{1}^{(i)}(t)\right),
   \label{eq:phi}
\end{equation}
where $\phi_{0}^{(i)}$ is the initial polar angle of the given robot at the moment $t=0$ and $\phi_{1}^{(i)}(t)$ is the polar angle at the moment $t$. For example, if $\rho$ is constant while $\phi$ changes considerably, it shows that the swarm rotates as a whole while slightly changing its geometry.

\subsection{2D correlation statistics}

\textit{Sixfold index $\psi_{6}$}~\cite{1992_Strandburg, 2021_Wang} represents spatial ordering, i.e., time-independent spatial correlations:
\begin{equation}
   \psi_{6}=\left\langle\frac{1}{N_{j}} \sum_{j^{\prime}} {\rm e}^{i 6 \theta_{j j^{\prime}}} \right \rangle_{\text{bulk}},
   \label{eq:Sixfold}
\end{equation}
where $N_{j}$ is the number of robots touching $j$'th robot and $\theta_{j j^{\prime}}$ is the angle between the position vectors of $j$'th and $j^{\prime}$'th robots. The operator $\langle\cdot\rangle_{\text {bulk}}$ denotes the average over all robots, excluding those placed near the border. This quantity reaches high values if the structure of the robots' packing resembles a hexagonal crystal (e.g., a close packing of cylindrical Type-I Swarmodroids), Figure~\ref{fig:char_demonstration}(d).

\begin{figure*}
    \centering
    \includegraphics[width=\textwidth]{Fig4.pdf}%
    \label{fig:Individual_dynamics}
    \caption{Properties of individual (a-d) Type-I (circular, self-rotating) and (e-f) Type-II (oval-shaped, self-propelled) Swarmodroids. (a) Vibration spectra of four different Type-I robots. (b) Vibration spectra of a single Type-I robot at different PWM levels from $10\%$ to $50\%$ with a $10\%$ step. (c) Angular velocity $\omega_{i}$ averaged over five realizations for each of seven Type-I Swarmodroids at $\mathrm{PWM} = 20\%$. (d) Angular velocities $\omega_{i}$ as functions of the PWM level for seven different Type-I Swarmodroids. The values are obtained by repeating the measurement five times for each robot, and the error bars denote the dispersion. (e) Linear velocity $v_{i}$ averaged over five realizations for each robot at $\mathrm{PWM} = 20\%$. (f) The same as Panel (d), but for linear velocities $v_{i}$ of seven Type-II Swarmodroids at different PWM levels }
\end{figure*}

\begin{figure}
    \centering
    \includegraphics[width=8.5cm]{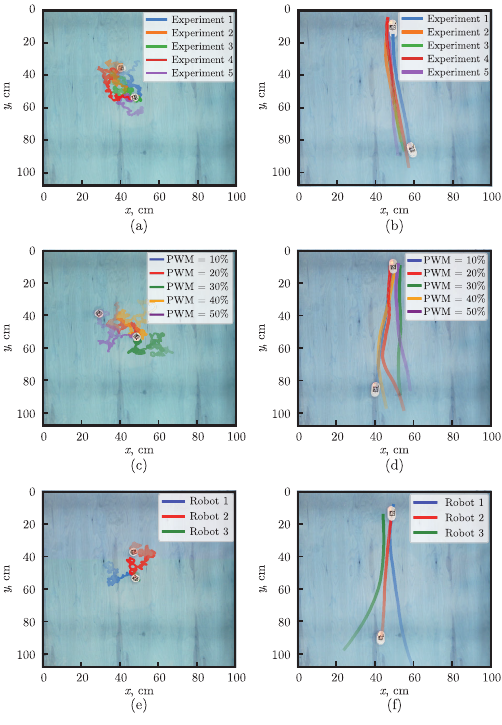}
   \label{Fig:trajectories}
   \caption{Motion trajectories for (a)-(c) Type-I (circular, self-rotating) and (d-f) Type-II (oval-shaped, self-propelled) Swarmodroids. (a),(d) The trajectories of the same robots moving at $\text{PWM}=20\%$ for different experiment realizations during (a) $60$~s (all experiments) and (d) $4$~s (the blue, orange, and green solid lines) and $5$~s (the red and purple solid lines). (b),(e) The trajectories of the same robots for single experiment realizations at different PWM levels. The experiment durations are (b) $60$~s for all ${\rm PWM}$ levels and (e) $5$~s for ${\rm PWM}=10\%$, $20\%$ (the blue and red solid lines) and $2$~s for ${\rm PWM}=30\%$, $40\%$, $50\%$ (the green, orange, and purple solid lines). (c),(f) The trajectories at $\text{PWM}=10\%$ for three different robots moving for (c) $60$~s and (f) $11$~s (Robot~$1$, the blue solid line), $6$~s (Robot~$2$, the red solid line), and $5$~s (Robot~$3$, the green solid line). The two images of the robot at each panel denote its initial (semi-transparent) and final (opaque) configurations in a single experiment.}
\end{figure}

\textit{Spatio-temporal correlation parameter $\tau_{*}$}~\cite{2007_Keys, 2021_Wang} reflects the time-correlated spatial dynamics of the robots. It is defined by the four-point susceptibility order parameter $\chi_{4}$ depending on the dynamical overlap function:
\begin{equation}
   Q(t, \tau ; a)=\frac{1}{N} \sum_{j=1}^{N} \Theta\left(a-\left|\hat{r}_{j}(t+\tau)-\hat{r}_{j}(t)\right|\right),
\end{equation}
where $a$ is the characteristic length chosen as the robot's radius, $t$ is the start timestamp, $\tau$ is the time from the start, and $\Theta$ is the Heaviside step function. The position vector of the $j$'th robot at the timestamp $t$ is given by
\begin{equation}
   \hat{r}_{j}(t)=x_{j}(t)\hat{x}+y_{j}(t)\hat{y},
\end{equation}
where $\hat{x}$ and $\hat{y}$ are unit vectors. The sustainability parameter $\chi_{4}(\tau ; a)$ can be evaluated as the variance of $Q(t, \tau ; a)$ over the time interval:
\begin{equation}
   \chi_{4}(\tau ; a)=N \operatorname{Var}_{t}(Q(t, \tau ; a)).
\end{equation}
Then, according to Ref.~\cite{2021_Wang},
\begin{equation}
   \tau^* = \text{max}_\tau \chi_{4}(\tau ; a)
   \label{eq:tau}
\end{equation}
is a characteristic trapping time for the robot around a given position, see Figure~\ref{fig:char_demonstration}(e).

\subsection{3D correlation statistics}

\textit{Two-dimensional pair correlation}~\cite{2010_Zhang} quantifies the probability per unit area (normalized by the area density $\rho$) of finding another robot at the location $(x,y)$ away from the reference robot, Figure~\ref{fig:char_demonstration}(f):
\begin{equation}
   g(x,y)=\frac{A}{N_{\rm total}}\left\langle\sum_{j \neq i} \delta\left[x \hat{x}_{i}+y \hat{y}_{i}-\left(\hat{r}_{i}-\hat{r}_{j}\right)\right]\right\rangle_{i},
   \label{eq:Correlation}
\end{equation}
where $\delta$ is pseudo-Dirac function ($\delta(0) = 1$ instead of $\infty$), $N_{\text{total}}$ is the total number of robots, $A$ is a scaling factor, $\hat{r}_{i}$ is the radius-vector of the $i$'th robot center, $\hat{x}_{i}$ and $\hat{y}_{i}$ are the transverse and longitudinal axes with respect to $\hat{r}_{i}$, and $\langle\ldots\rangle_{i}$ represents averaging over all robots. The physical meaning of this quality is the following. If the centers of robots cannot locate at a certain distance from each other (e.g., at a distance closer than the robot's diameter), $g(x,y)$ will tend to zero, while at the characteristic distances between robots packed in typical clusters, crystalline lattices, etc., the values of $g(x,y)$ will be finite, as shown in Figure~\ref{fig:char_demonstration}(f). A similar picture has been experimentally demonstrated for self-rotating robots~\cite{2021_Dmitriev}.

\textit{Orientation correlation function}~\cite{2010_Zhang} reflects the orientation dependencies between the robots: 
\begin{equation}
   C_{\theta}(x, y)=\left\langle\left(\hat{y}_{i} \cdot \hat{y}_{j}\right) \delta\left[x \hat{x}_{i}+y \hat{y}_{i}-\left(\hat{r}_{i}-\hat{r}_{j}\right)\right]\right\rangle_{ij},
\end{equation}
where $\langle\ldots\rangle_{ij}$ represents averaging over all possible pairs. According to the formula, the parameter $C_{\theta}$ tends to have higher values when robots at the location $(x,y)$ are oriented in the same direction as the reference robot. In the case of robots without circular symmetry, such as Type-II Swarmodroids, this quantity characterizes the spatial alignment of the robots.

\textit{Velocity correlation function}~\cite{2010_Zhang} allows to capture the velocity dependencies between the robots:
\begin{equation}
   C_{v}(x, y)=\frac{\left\langle\left(\hat{v_{i}} \cdot \hat{v_{j}}\right) \delta\left[x \hat{x}_{i}+y \hat{y}_{i}-\left(\hat{r_{i}}-\hat{r_{j}}\right)\right]\right\rangle_{i j}}{\left\langle\hat{v_{i}} \cdot \hat{v_{i}}\right\rangle_{i}},
\end{equation}
where $\hat{v_{i}}$ is the velocity vector of the $i$'th robot. Such a quantity will reach its maximal value when the velocity directions of all robots are aligned, and can be useful in visualizing flocking and other alignment phenomena.

\section{Dynamics of individual robots}
\label{sec:Individual}

To engineer the collective behavior of robotic swarms or study their physics, one needs to have knowledge of the parameters corresponding to individual robots. To this end, we performed a detailed characterization of individual Swarmodroids addressing their angular (for Type-I) and linear (for Type-II) velocities, vibration spectra, and evolution of these parameters upon changing the PWM duty cycle.

To measure the vibration spectra of robots, we use an IMVVP-4200 accelerometer working at the sampling frequency $f_{\rm s}=10$~kHz. To ensure a rigid connection, the accelerometer is fastened to a modified cap with a hole having the same shape as the accelerometer. During vibration frequency measurements, Swarmodroids are attached to the table with the help of two-sided adhesive tape to limit the magnitude of their oscillations and improve the quality of the measurements. We obtain the vibration amplitude sampled over time using the LabVIEW software package. Then, we apply the Fourier transform to process the extracted time series and evaluate the vibration spectrum of each robot.

Figure~\ref{fig:Individual_dynamics}(a) demonstrates the vibration spectra of four different Type-I Swarmodroids, all working at $\mathrm{PWM}=20\%$. The spectra feature the presence of a pronounced peak corresponding to the main mode with a frequency around $f_{0} \approx 250$~Hz for $\mathrm{PWM}=20\%$ surrounded by significantly lower peaks of other modes. The main frequency $f_{0}$ remains nearly the same for all the robots considered, while the structure of the other peaks may fluctuate considerably. Figure~\ref{fig:Individual_dynamics}(b) demonstrates such spectra for a single Type-I Swarmodroid working at different PWM duty cycles from $10\%$ to $50\%$, respectively. When the PWM duty cycle is increased, the frequency of this main mode changes linearly from $f_{0} \approx 180$~Hz for $\mathrm{PWM}=10\%$ to higher values, up to $f_{0} \approx 385$~Hz for $\mathrm{PWM}=50\%$. Thus, the swarm can be approximately described with a single characteristic vibration frequency $f_{0}$, which depends linearly on the PWM duty cycle.

\begin{table*}
\def\arraystretch{1.5}
\centering
\begin{tabular}{p{16mm} p{4mm} p{11mm} p{9mm} p{16mm} p{13mm} p{12mm} p{16mm} p{28mm} p{9mm} p{4mm}}
\hline
Platform	& Year	& Size, cm	& Mass, g	& Linear motion velocity, cm/s	& Rotation freq., rad/s	& Processing software	& Recognition technology	& Devices for robot control & Price, USD \\
\hline
\multicolumn{10}{c}{\rule{0pt}{10pt}\textbf{Wheeled robots}} \\
\hline
Alice                & 2000	& 2.2    & 5	& 4	    & -	        & custom	& LED               & Sensors              & - \\
Jasmine-III          & 2005	& 3      & -	& 30	& -	        & custom	& -               & Remote speed control & 130 \\
AMiR                 & 2009	& 6.5    & -	& 8.6	& -	        & WhyCon	& Markers           & Remote speed control & 78 \\
e-Puck 2             & 2009	& 7      & 150	& 15	& -	    & IRIDA	    & Markers           & Sensors              & 1200 \\
Elisa-3              & 2013	& 5      & 39	& 60	& -	        & SwisTrack	& IR emitters	    & Remote speed control & 390 \\
Colias               & 2014	& 4      & 28	& 35	& -	        & custom	& Markers           & Sensors              & 30 \\
\hline
\multicolumn{10}{c}{\rule{0pt}{10pt}\textbf{Bristle-bots}} \\
\hline
Hexbug               & 2007	& 4.3    & 7	& 40	& -	        & custom	& Colored spots	    & None       	       & 5 \\
BBots                & 2012	& 7.92   & 15.5	& 20	& 3         & -	    & -	            & None       	       & - \\
Kilobot              & 2012	& 3.3    & 16	& 1     & 0.8	    & trackpy	& Shape      	    & Remote speed control & 14 \\
BOBbots              & 2021	& 6      & 60	& 4.8	& 1.9	    & -	    & -	            & Remote speed control & - \\
SurferBot            & 2022	& 5      & 2.6	& 10	& -	        & imaqtool	& Colored dots	    & None       	       & - \\
{SimoBot}               & {2022}	&  {2}  &  {4.76} &   {4}   &  {-}	&  {custom}	&        {Markers}        &  {Remote speed control} & {4.7}  \\
Magbot               & 2024	& 2      & -	& 2     & 0.5--13.6	& custom	& LED               & Remote speed control & - \\
{MARSBot}               & {2024}	&  {4.7}  &  {24} &   {6.813}   &  {-}	&  {-}	&        {-}        &  {AR steering (headset)} & {-}  \\
Swarmodroid Type-I   & 2025	& 5      & 21	& -	    & 6.3--12.6	& AMPy	    & Markers  & Remote speed control & 11 \\
Swarmodroid Type-II	& 2025	& 8.5    & 23	& 5--40	& -	        & AMPy  	& Markers  & Remote speed control & 11 \\
\hline
\end{tabular}
   \caption{Comparison of several wheeled robots~\cite{2000_Caprari_Alice, jasmine, 2009_Arvin_AMIR, gtronic_e_Puck, elisa3, 2014_Arvin_Colias} and bristle-bots~\cite{2018_Deblais, 2013_Giomi, 2012_Rubenstein_Kilobot, 2021_Li, 2022_Rhee_Surferbot, 2022_Zhang_SimoBot, 2024_Wang, 2024_Fath}, including Swarmodroid 1.0. Column ``Size'' contains the largest dimension of the robot. ``Linear motion velocity'' and ``Rotation frequency'' show the maximal values of the respective parameters. Columns ``Processing software'' and ``Recognition technology'' highlight the tools allowing to track the robot position. Column ``Devices for robot control'' specifies whether the robots are equipped with sensors (e.g., infrared or ultrasound) for interaction or orientation in surrounding space, or devices that only allow remote control over their speed, which also facilitate simultaneous activation of all robots in the swarm. Column ``Price'' lists the cost of purchase or assembly of a single robot for the respective platform in US dollars as of 2023, if available.}
   \label{table:Platform}
\end{table*}

The angular velocities $\omega_i$ of seven Type-I Swarmodroids experimentally measured at different PWM levels are shown in Fig.~\ref{fig:Individual_dynamics}(c). It is seen that the velocities grow monotonically for all considered robots upon increasing the PWM duty cycle. However, a certain degree of dispersion of the angular velocity values is observed at low ${\rm PWM=10}\%$ and $\mathrm{PWM}=20\%$, which becomes considerable for larger PWM values. Figure~\ref{fig:Individual_dynamics}(e) demonstrates velocities $v_i$ of seven Type-II Swarmodroids in a similar fashion. Similarly to the self-rotating robots shown in Fig.~\ref{fig:Individual_dynamics}(c), self-propelled ones demonstrate monotonic growth of velocities with increasing the PWM level. The velocities $v_i$ and the angular velocities $\omega_i$ for $\mathrm{PWM}=20\%$ are shown in Figure~\ref{fig:Individual_dynamics}(d,f), respectively, to illustrate the stability of these parameters. {The maximum linear velocity fluctuation after averaging over five different realizations is approximately $\pm 2.5$~cm/s, while the maximum deviation from the mean angular velocity is approximately $\pm 1.5$ revolutions per second.}

{To study the properties of Swarmodroid trajectories, we perform several measurements of individual robots motion shown in Figure~\ref{Fig:trajectories}. For Type-I Swarmodroids shown in Fig.~\ref{Fig:trajectories}(a-c), all trajectories were captured for $60$~s. It is seen that along with a self-rotation, some displacement of robots is observed resembling a random walk. The shape of the trajectory differs in experiments with the same robot, demonstrating that it is related to various imperfections in the surface at which the robot moves as well as in the robot construction instead of some systematic properties, Fig.~\ref{Fig:trajectories}(a). Moreover, Fig.~\ref{Fig:trajectories}(b) demonstrates that this characteristic displacement is independent of the PWM value, i.e., angular velocity of the robot. Finally, different robots demonstrate qualitatively similar displacement trajectories, as shown in Fig.~\ref{Fig:trajectories}(c). For Type-II Swarmodroids, the trajectories are nearly straight, as seen in Fig.~\ref{Fig:trajectories}(d-f). While the variance of trajectory between different experiments for a single robot in Fig.~\ref{Fig:trajectories}(d) is less pronounced, it is seen that the trajectory depends on the PWM value, Fig.~\ref{Fig:trajectories}(e). Finally, different robots may possess some chiral contributions, either CW- or CCW-, as shown in Fig.~\ref{Fig:trajectories}(f).}

\section*{Outlook}
\label{sec:Outlook}

In the present paper, we introduce an open-source Swarmodroid platform featuring bristle-bots with a remote IR control, a set of 3D printed plastic parts to reconfigure them for different application scenarios, and a software package capable of automatic extraction of various quantities characterizing the behavior of the swarm. The developed robot design offers a certain degree of control over its motion velocity by setting the vibration motor power via the PWM duty cycle in response to commands received from the IR remote control. As demonstrated by studies of individual robots, they can be described by a characteristic vibration frequency $f_{0}$ that slightly deviates between different robots and increases linearly from $f_{0} \approx 180$~Hz to $f_{0} \approx 380$~Hz with an increase in the pulse modulation width of the vibration motor from $\mathrm{PWM}=10\%$ to $\mathrm{PWM}=50\%$. The self-rotation angular velocities of Type-I Swarmodroids and motion velocities of the self-propelled Type-II Swarmodroids grow monotonically upon such an increase in PWM as well, which allows one to control the dynamics of the swarm on the go by changing the PWM duty cycles with the help of an IR remote.

\begin{figure}
    \includegraphics[width=8.5cm]{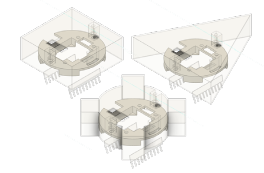}
    \label{Fig:body_design}
    \caption{Renders of various Swarmodroid body designs that can be implemented by replacing the upper cap only and may result in different collective behaviors. }
\end{figure}

Table~\ref{table:Platform} summarizes the key characteristics of several swarm robotic platforms, including wheeled robots and bristle-bots. The Swarmodroid is characterized by high robot speed and tunability (including the ability to change between linear and rotating motions), at the same time making large swarms feasible. The latter is facilitated by the open-source distribution model, the low cost of a single robot, and the availability of ready-to-use tracking software. Therefore, the proposed platform can be effectively applied to perform experimental studies in various areas of many-body physics, biology, transportation, and engineering applications.
\begin{itemize}
   \item In physics, such robotic swarms can be used as models for various phenomena~\cite{2022_DiBari}, {to experimentally demonstrate} novel theoretical predictions that cannot yet be implemented in natural materials~\cite{2025_Veenstra}, or even as a source of {new} experimental data~\cite{2022_Boudet}. Regarding our platform, we propose to tackle the problems that require variable velocity or specific shapes of robotic bodies, {like those illustrated in Figure~\ref{Fig:body_design}. For example, Swarmodroids were recently applied to study swarms of teardrop-shaped robots demonstrating the formation of clusters that resemble micelles in surfactant solutions~\cite{2025_Molodtsova}. Moreover, one can apply the proposed platform to study different patterns of self-organization~\cite{2017_Lenz} with aim to realize shape-morphing matter~\cite{2022_Xia, 2021_Chvykov, 2015_Romanishin, 2018_Daudelin, 2024_Wang, 2024_Saintyves}, e.g., by implementing time-dependent profiles of the PWM duty cycle or tuning the shape of Swarmodroid caps.}
   
   \item {In biology}, robotic swarms can be applied to mimic the behavior strategies of various biological systems, such as worm blobs~\cite{2021_Ozkan_Aydin}, magnetotactic bacteria~\cite{2021_Sepulveda}, {cell collectives~\cite{2024_Pan},} or insect colonies~\cite{2009_Garnier, 2014_Werfel, 2022_Prasath}. {In this sense, it looks promising to incorporate metallic parts and additional magnets in Swarmodroid caps to study magnetic interactions between robots~\cite{2021_Li, 2024_Wang, 2024_Saintyves}, and consider different complex shapes of their caps to further delve into geometry-mediated self-organization based on differential adhesion~\cite{2024_Pan}.}
   
   \item {In pedestrian dynamics, many effects} are modeled with the help of particle swarms~\cite{2019_Nicolas, 2022_Echeverria_Huarte}, including hydrodynamic approaches~\cite{2018_Filella, 2019_Bain}. In this light, our platform can be used to {experimentally} consider phenomena such as jamming of pedestrians in narrow exits~\cite{2014_Karamouzas, 2019_Haghani}, emulate interactions governed by simple rules~\cite{2021_Murakami}, and study the formation of collective structures~\cite{2013_Silverberg, 2025_Gu}. Besides, such robotic swarms can be applied to experimentally implement various simplified traffic models~\cite{2002_Nagatani, 2019_Barois}.
   
   \item {In engineering, such bristle-bots can be applied to perform the inspection of pipes~\cite{2008_Wang, 2014_Becker},} obstructions~\cite{2003_Wang}, hazardous environments, space infrastructure~\cite{2022_Haghighat}, and geological objects by swarms of robots equipped with sensors and transmitters in cases where required robot sizes are strictly limited, or when the robots are likely to be destroyed, and minimizing their cost is important. {For example, a bristle-bot carrying a camera has been introduced~\cite{2024_Fath} for monitoring of narrow spaces and applied in a heterogeneous robot collective~\cite{2025_Fath}. Bristle-bots were recently demonstrated to navigate the surface of water~\cite{2022_Karavaev, 2022_Rhee_Surferbot} which can be used to perform its monitoring. Moreover, one can consider an implementation} of universal grips using the jamming transition that occurs in dense swarms. Such devices were demonstrated considering jamming in a passive granular medium~\cite{2010_Brown}. One can start with Swarmodroids placed in a flexible barrier similar to those of Refs.~\cite{2018_Deblais, 2021_Boudet}, but at higher densities compared to the mentioned papers.

\end{itemize}

Finally, we outline the potential directions for further development of the Swarmodroid platform.
\begin{itemize}
   
   \item {The most recent bristle-bots feature two degree of freedom steering, implemented either by incorporating two vibration motors~\cite{2012_Rubenstein_Kilobot}, using two-frequency driving to excite different vibration modes~\cite{2017_Turkmen, 2025_Hao}, or by changing the rotation direction of the motor~\cite{2022_Zhang_SimoBot}. Such a capability is essential for single-robot applications as well as for introducing more complex swarm control paradigms linked to machine learning~\cite{2023_Zion} or phototactic behavior~\cite{2021_Wang}.}
   
   \item {Various sensors can be introduced to increase the capabilities of single Swarmodroids and allow for more complex swarming behaviors. For example, several realizations of bristle-bots with cameras have recently been introduced~\cite{2020_Iyer, 2024_Fath, 2025_Hao}. Moreover, temperature and humidity sensors~\cite{2025_Fath} have also been attached to bristle-bots, and the use of light sensing is quite common~\cite{2012_Rubenstein_Kilobot, 2021_Wang, 2023_Siebers}. However, due to the limited resources of the ATTiny 13 microcontroller, this will require its substitution with a more powerful alternative, for example, ATMega microcontrollers.}
   
   \item {The introduction of wireless charging functionality will substantially increase the convenience of robotic charging, which is important for the accumulation of large experimental datasets, such as $300$ identical experiments performed with Swarmodroids in~\cite{2023_Buzakov} considering the formation of polycrystalline clusters by robots moving in a parabolic potential. Although there are different demonstrations of wireless energy transmission to swarms of moving objects, ranging from powering submillimeter microsystems with resonant inductive power transfer at frequencies $3.5..3.8$~kHz~\cite{2020_Bandari} to $5$~GHz radiative power transmission to a centimeter-scale flapping-wing aerial vehicle~\cite{2021_Ozaki}, the most common (and, thus, the most accessible for production) wireless power transfer standard is Qi~\cite{2010_VanWageningen} working at the frequencies of $100..200$~kHz that was applied for developing a large-area charger for compact robots~\cite{2018_Arvin} as well as in BOBbots~\cite{2021_Li}. Incorporating Qi receiving coils in Swarmodroids looks promising, considering that this standard has recently been applied to construct a rechargeable AA battery with a curved receiving coil~\cite{2025_Dmitriev}, demonstrating its suitability for further miniaturization.}

\end{itemize}

We encourage all members of the community to introduce their ideas and develop modifications of the proposed Swarmodroid platform.

\section*{Acknowledgements}
The authors acknowledge valuable discussions with Anton Souslov, Dmitry Filonov, Denis Butusov, Evgenii Svechnikov, and Egor Kretov. 

\section*{Author contributions}
Alexey Dmitriev designed the printed circuit boards and developed the firmware. Vadim Porvatov and Mikhail Buzakov developed the AMPy package. Alina Rozenblit and Anastasia Molodtsova designed Swarmodroid bodies and optimized bristles. Daria Sennikova, Vyacheslav Smirnov, Mikhail Buzakov, and Timur Karimov performed studies of individual Swarmodroids. Oleg Burmistrov measured the discharge characteristics of the robots. Ekaterina Puhtina, Alina Rozenblit, Alexey Dmitriev, and Oleg Burmistrov soldered PCBs and assembled the robots. Nikita Olekhno put forward the idea and supervised the project. All authors contributed to the preparation of the manuscript, data processing, and discussion of the results.

\section*{Code availability}
The source code for the Swarmodroid firmware is available at~\url{https://github.com/swarmtronics/swarmodroid.firmware}. The source code of the AMPy package for video data processing is available at~\url{https://github.com/swarmtronics/ampy}. The electric circuit diagram and the printed circuit board layouts of the Swarmodroid are available at~\url{https://github.com/swarmtronics/swarmodroid.pcb}.


\bibliographystyle{unsrt}

\clearpage
\includepdf[pages={1}]{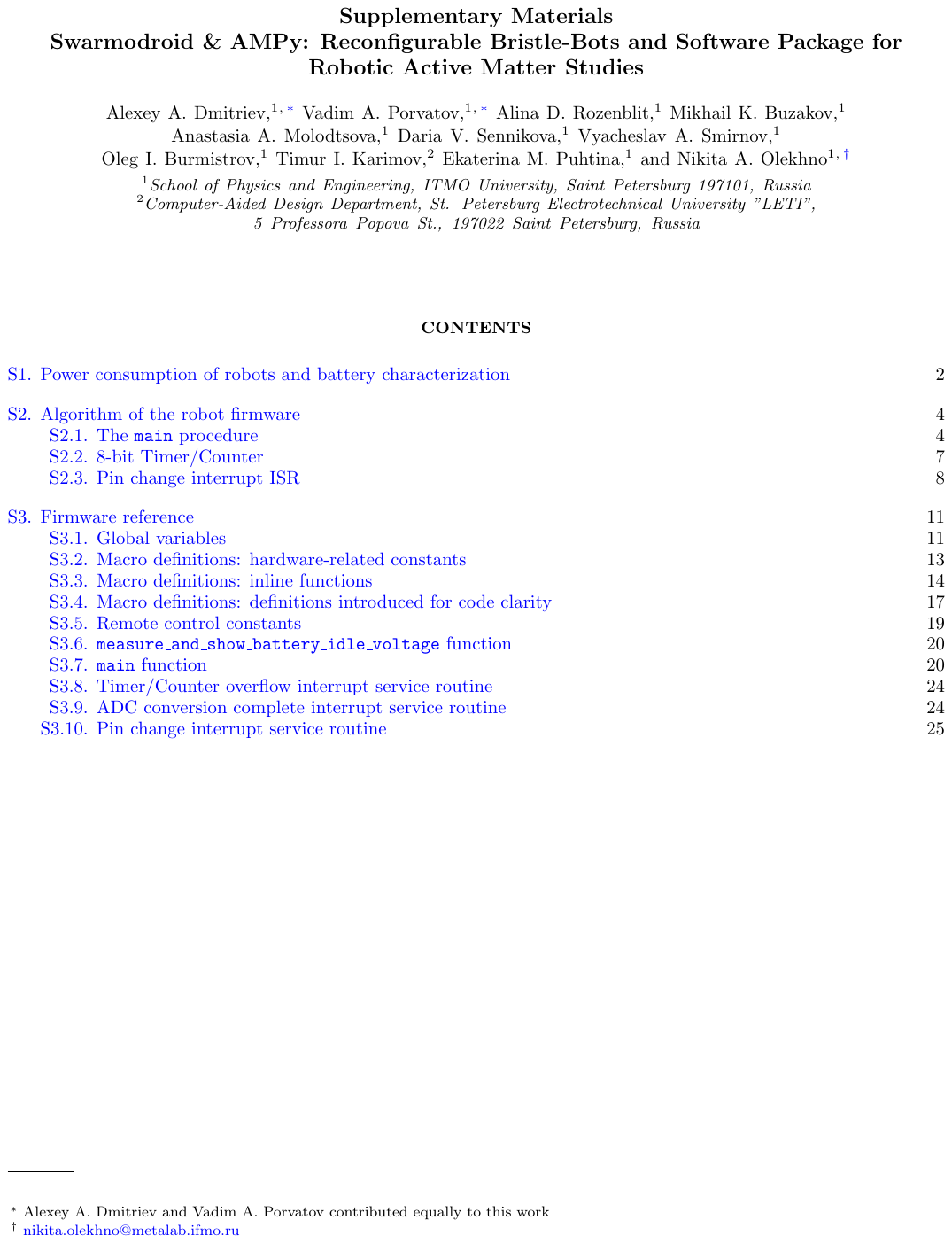}
\clearpage
\includepdf[pages={2}]{SupplementaryPreCompiled.pdf}
\clearpage
\includepdf[pages={3}]{SupplementaryPreCompiled.pdf}
\clearpage
\includepdf[pages={4}]{SupplementaryPreCompiled.pdf}
\clearpage
\includepdf[pages={5}]{SupplementaryPreCompiled.pdf}
\clearpage
\includepdf[pages={6}]{SupplementaryPreCompiled.pdf}
\clearpage
\includepdf[pages={7}]{SupplementaryPreCompiled.pdf}
\clearpage
\includepdf[pages={8}]{SupplementaryPreCompiled.pdf}
\clearpage
\includepdf[pages={9}]{SupplementaryPreCompiled.pdf}
\clearpage
\includepdf[pages={10}]{SupplementaryPreCompiled.pdf}
\clearpage
\includepdf[pages={11}]{SupplementaryPreCompiled.pdf}
\clearpage
\includepdf[pages={12}]{SupplementaryPreCompiled.pdf}
\clearpage
\includepdf[pages={13}]{SupplementaryPreCompiled.pdf}
\clearpage
\includepdf[pages={14}]{SupplementaryPreCompiled.pdf}
\clearpage
\includepdf[pages={15}]{SupplementaryPreCompiled.pdf}
\clearpage
\includepdf[pages={16}]{SupplementaryPreCompiled.pdf}
\clearpage
\includepdf[pages={17}]{SupplementaryPreCompiled.pdf}
\clearpage
\includepdf[pages={18}]{SupplementaryPreCompiled.pdf}
\clearpage
\includepdf[pages={19}]{SupplementaryPreCompiled.pdf}
\clearpage
\includepdf[pages={20}]{SupplementaryPreCompiled.pdf}
\clearpage
\includepdf[pages={21}]{SupplementaryPreCompiled.pdf}
\clearpage
\includepdf[pages={22}]{SupplementaryPreCompiled.pdf}
\clearpage
\includepdf[pages={23}]{SupplementaryPreCompiled.pdf}
\clearpage
\includepdf[pages={24}]{SupplementaryPreCompiled.pdf}
\clearpage
\includepdf[pages={25}]{SupplementaryPreCompiled.pdf}
\clearpage
\includepdf[pages={26}]{SupplementaryPreCompiled.pdf}
\clearpage
\includepdf[pages={27}]{SupplementaryPreCompiled.pdf}
\clearpage
\includepdf[pages={28}]{SupplementaryPreCompiled.pdf}
\clearpage
\includepdf[pages={29}]{SupplementaryPreCompiled.pdf}

\end{document}